\documentclass[prl,amsmath,amssymb,twocolumn,superscriptaddress]{revtex4}

\usepackage{amsmath,amssymb}
\usepackage[usenames]{color}
\usepackage{amssymb}
\usepackage{dsfont}
\usepackage{grffile}
\usepackage[pdftex]{graphicx}
\usepackage{amsmath, amstext, amssymb, amsfonts, amsxtra}
\usepackage{textcomp}
\usepackage{xspace}
\DeclareSymbolFontAlphabet{\amsmathbb}{AMSb}

\newcommand{\cur}{\mathcal{J}}
\newcommand{\curop}{\hat{j}}
\newcommand{\uw}{\uparrow}
\newcommand{\dw}{\downarrow}
\newcommand{\Rec}{\mathcal{R}}
\newcommand{\Con}{\mathcal{C}}
\newcommand{\ket}[1]{|\!\! #1 \rangle}
\newcommand{\bra}[1]{\langle #1 \!\!|}

\newcommand{\ave}[1]{\langle #1 \rangle }

\newcommand{\rhop}{\hat{\rho}}

\newcommand{\Hop}{\hat{H}}
\newcommand{\sop}{\hat{\sigma}}
\newcommand{\Dop}{\mathcal{D}}
\newcommand{\tr}{{\rm tr}}

\newcommand{\im}{{\rm i}}

\newcommand{\la}{\lambda}
\newcommand{\Lop}{\mathcal{L}}

\newcommand{\Oop}{\hat{\Omega}}
\newcommand{\zop}{\hat{\zeta}}

\newcommand{\sutd}{EPD Pillar, Singapore University of Technology and Design, 8 Somapah Road, 487372 Singapore}
\newcommand{\como}{Center for Nonlinear and Complex Systems, Dipartimento di Scienza e Alta Tecnologia, Universit\`a degli Studi dell``Insubria, via Valleggio 11, 22100 Como, Italy}
\newcommand{\infn}{Istituto Nazionale di Fisica Nucleare, Sezione di Milano, via Celoria 16, 20133 Milano, Italy}
\newcommand{\nest}{NEST, Istituto Nanoscienze-CNR, I-56126 Pisa, Italy}
\newcommand{\brasil}{Departamento de F\'{\i}sica--Instituto de Ci\^{e}ncias Exatas, Universidade Federal de Minas Gerais, CP 702, 30.161-970 Belo Horizonte MG, Brazil}
\newcommand{\riogrande}{International Institute of Physics, Federal University of Rio Grande do Norte, Natal, Brazil}

\begin{document}

\title{Perfect diode in quantum spin chains}

\author{Vinitha Balachandran}
\affiliation{\sutd}
\author{Giuliano Benenti}
\affiliation{\como}
\affiliation{\infn}
\affiliation{\nest}
\author{Emmanuel Pereira}
\affiliation{\brasil}
\author{Giulio Casati}
\affiliation{\como}
\affiliation{\riogrande}
\author{Dario Poletti}
\affiliation{\sutd}

\begin{abstract}
We study the rectification of spin current in $XXZ$ chains segmented in two parts, each with a different anisotropy parameter. Using exact diagonalization and a matrix product states algorithm we find that a large rectification (of the order of $10^4$) is attainable even using a short chain of $N=8$ spins, when one half of the chain is gapless while the other has large enough anisotropy.
We present evidence of diffusive transport when the current is driven in one direction and of a transition to an insulating behavior of the system when driven in the opposite direction, leading to a perfect diode in the thermodynamic limit.
The above results are explained in terms of matching of spectrum of magnon excitations between the two halves of the chain. 
\end{abstract}


\maketitle

{\it Introduction}:
A key challenging problem for modern physics is understanding and controlling transport properties of many-body quantum systems. Important results have been obtained for boundary driven systems, i.e. for systems coupled to baths at their extremities.
Numerical and analytical results have shown that, by tuning interactions,
disorder, noise, external fields and coupling to the baths one can modify the transport properties and access different regimes, including ballistic, diffusive, sub-diffusive, and insulating behavior \cite{Prosen2011, Prosen2011b, Prosen2014, Karevski2013,Popkov2013, Znidaric,Prosen2016,NDC,BenentiZnidaric2009,GuoPoletti2017}.

Gaining a deeper understanding and control of transport properties at the nanoscale can lead to important technological advances. For instance, many-body nonlinear dynamics might be exploited to design nonlinear devices like heat diodes and transistors \cite{TerraneoCasati2002,LiCasati2004,Baowen2012,Benenti2016}. Quantum aspects in this quest are also attracting growing interest \cite{WerlangValente2014,LifaLi2009,Landi2014,Landi_many,ArracheaAligia2009}. However, so far the relevance of phase transitions \cite{MitraMillis2006, Diehl2008,Diehl2010,DallaTorre2010,GuoPoletti2016} for rectification has not been explored.

In this work we study a spin chain segmented in two parts, each with different anisotropy.
We couple the chain to two different magnetization
baths at its edges and study the spin current in forward or reverse bias.
Even for relatively small system sizes ($N=8-10$ spins), we find remarkable
rectification (of the order of $10^4$).
We show that the rectification is due to a mismatch in the
spectrum of magnon excitations
of one partition of the chain compared to the other,
a mismatch which only occurs in reverse bias. The mismatch is most prominent
when one half of the chain is noninteracting (\emph{i.e.} XX model),
while the other has strong interactions
(\emph{i.e.} XXZ model in the gapped phase with large enough anisotropy).
Our numerical and analytical results strongly indicate, in reverse bias,
a transition to an insulating behavior beyond a critical value of the anisotropy,
leading to perfect rectification in the thermodynamic limit.

{\it Model}: We consider a bipartite spin-$1/2$ chain described
by the Heisenberg XXZ Hamiltonian
\begin{align}\label{ham}
  \Hop=\sum_{n=1}^{N-1}\left[J_n(\sop_{n}^{x}\sop_{n+1}^{x}+\sop_{n}^{y}\sop_{n+1}^{y})+\Delta_{n}\sop_{n}^{z}\sop_{n+1}^{z}\right],
\end{align}
where $N$ (which we take to be even) is the total number of sites in the chain, $\sop_{n}^{\alpha}$, with $\alpha=x,y,z$ are the Pauli matrices for the $n$-th spin, $J_n$ and $\Delta_n$ are the strengths of the XX tunneling and the ZZ coupling, the anisotropy parameter being $\Delta_n/J_n$.
The parameters $J_n$ and $\Delta_n$ are chosen such that $\Delta_{n}=\Delta_{L}$ and $J_{n}=J_{L}$ when $n < N/2$ and $ \Delta_{n}=\Delta_{R}$ and $J_{n}=J_{R}$ for $n > N/2$.
At the junction between the two halves of the chain $\Delta_{N/2}=\Delta_{M}$ and $J_{N/2}=J_{M}$.

The chain is coupled to two different 
baths at
its extremities, each tending to impose a particular magnetization.
The evolution of the density operator $\rhop$ obeys the
Lindblad master equation \cite{Lindblad,GoriniSudarshan}
\begin{align}\label{mastereq}
  \frac{d\rhop}{d t}=\Lop(\rhop)=-\frac{\im}{\hbar}[\Hop,\rhop]+\sum_{n=1,N}\Dop_n(\rhop),
\end{align}
where $\hbar$ is the (reduced) Planck constant, $\Lop$ is the Lindbladian superoperator and the dissipator $\Dop_n$ on sites $n=1,N$ is given by
\begin{align}
\Dop_n(\rhop)=&\gamma \left[ \la_n \left(\sop^+_n \rhop \sop^-_n - 1/2 \left\{\sop^-_n\sop^+_n,\rhop\right\}  \right)   \right.   \nonumber \\
&+\left.  (1-\la_n) \left(      \sop^-_n \rhop \sop^+_n - 1/2 \left\{\sop^+_n\sop^-_n,\rhop\right\} \right)   \right].
\end{align}
Here, $\sop^+_n=(\sop^-_n)^{\dagger}=(\sop^x_n+\mathcal{\im} \sop^y_n)/2$, the parameter $\gamma$ is the intensity of the coupling to the baths while the spin magnetization imposed by the baths is set by $\la_n$. We use $\la_n\in [0,0.5]$ and in most cases we focus on $\la_1$ and $\la_N$ to be equal to $0$ or $0.5$, i.e. on one side of the chain the bath tends to set the spins to be pointing down ($\ket{\dw}_n\bra{\dw}$ for $\la_n=0$) or to be in an equal mixture of up and down spins ($[\ket{\dw}_n\bra{\dw} + \ket{\uw}_n\bra{\uw}]/2$ for $\la_n=0.5$). The first case corresponds to the coupling to a magnet, acting as a reservoir for magnetization \cite{loss}, the second to a high temperature bath.
 The imbalance $\lambda_1-\lambda_N$ imposed by the baths generates a spin current.
For the `forward bias' $\la_1>\la_N$ the spin current flows from left to right, while for the `reverse bias' $\la_1<\la_N$ the current flows from right to left.

The spin current $\cur$ is defined via the continuity equation for the local observable $\sop_n^z$ which gives $\cur={\rm tr}(\curop_n \rhop_{ss})$, where
$\curop_n=2J_n\left(\sop^x_n\sop^y_{n+1}-\sop^y_n\sop^x_{n+1}\right)/\hbar$
and $\rhop_{ss}$ is the steady state, $\Lop(\rhop_{ss})=0$. We refer to the forward(reverse) spin current as
$\cur_f$($\cur_r$).
To measure the spin current rectification, we calculate the
rectification coefficient
 \begin{equation}\label{rect2}
  \Rec=-\frac{\cur_{f}}{\cur_{r}}.
 \end{equation}
In absence of rectification $\Rec=1$ while $\Rec$ tends to infinity for a perfect
diode.

To observe a sharp signature of the rectification we also study the current contrast \cite{Landi2014, Landi_many}
 \begin{equation}\label{contrast}
  \Con=\left|\frac{\cur_f+\cur_r}{\cur_f-\cur_r}\right|.
 \end{equation}
Note that the contrast $\Con=0$ when there is no rectification and takes the value $\Con=1$ for a perfect diode.

\begin{figure}
\includegraphics[width=\columnwidth]{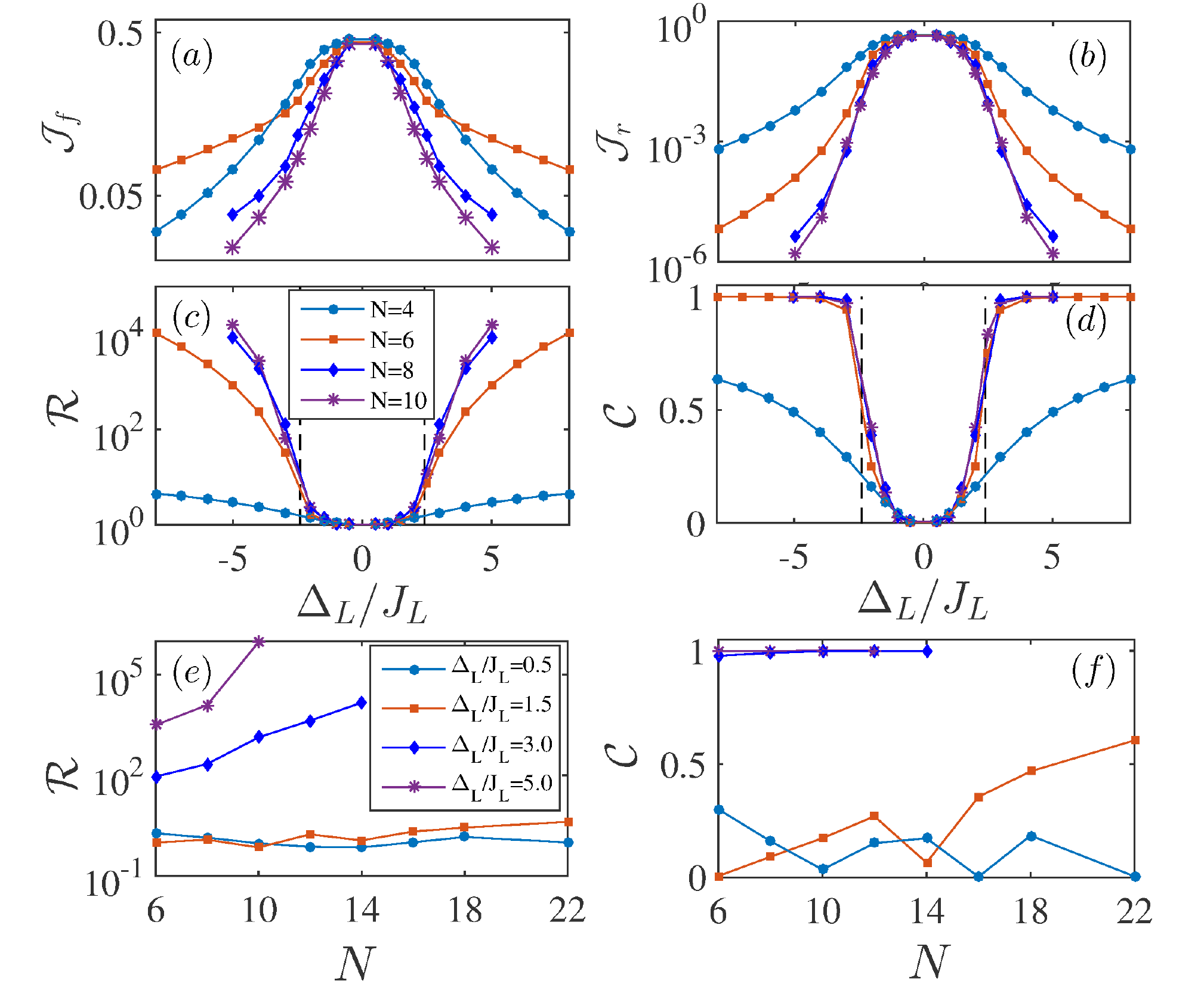}
\caption{(color online) (a-d) We show the forward current $\cur_f$ (a), the reverse current $\cur_r$ (b), the rectification coefficient $\Rec$ (c), and the contrast $\Con$ (d) as functions of the anisotropy $\Delta_L/J_L$ for different chain lengths $N$. The current in reverse bias is significantly lower than in forward bias resulting in large rectification and contrast.
(e,f) Rectification $\Rec$ (e) and contrast $\Con$ (f) as functions of the system size for different values of the anisotropy $\Delta_L/J_L$. The rectification $\Rec$ increases significantly with the system size for large enough anisotropy $\Delta_L/J_L$.
The other parameter values are $\Delta_R=\Delta_M=0$, $J_L=J_R$, $J_M=J_L$ and $\gamma=J_L/\hbar$ for (a-d), while $J_M=0.1 J_L$ and $\gamma=0.1 J_L/\hbar$ for (e-f). The black dashed lines in Fig.~\ref{fig:Fig1}(c-d) indicate $\Delta_L/J_L=1+\sqrt{2}$. }
\label{fig:Fig1}
\end{figure}

{\it Results}:
We study spin rectification in the above system by exact diagonalization for small chains and by a matrix product states algorithm for the evolution of the density matrices for larger systems \cite{VerstraeteMung,Schollwock2011}.
In most of our simulations we set $J_L=J_R$, $\Delta_R=\Delta_M=0$ (no $ZZ$ coupling on the right part of the chain and at the left-right interface) and  vary the anisotropy on the left portion of the chain by changing the parameter $\Delta_L$.
We will then later discuss the case $J_L\ne J_R$ and $\Delta_R\ne 0$.

In Fig.~\ref{fig:Fig1}(a,b) we show the forward and reverse-bias currents, $\cur_f$ and $\cur_r$. Both currents decrease as the
interactions (anisotropy $|\Delta_L/J_L|$) increase, however we can notice a marked difference in the magnitude of the forward and reverse currents at a large enough anisotropy. Hence we study both the rectification $\Rec$ [Fig.~\ref{fig:Fig1}(c)] and the contrast $\Con$ [Fig.~\ref{fig:Fig1}(d)]. We notice that the rectification sharply increases for $|\Delta_L/J_L| \approx 1+\sqrt{2}$ (this precise number will be justified later) and, already for chains of size $N=8-10$, it can reach values of $10^4$ for an anisotropy $|\Delta_L/J_L|\approx 4$.
The increase in the rectification becomes more pronounced for longer chains, suggesting the possible occurrence of a transition in the thermodynamic limit. To visualize this effect more clearly we study the contrast $\Con$, shown in Fig.~\ref{fig:Fig1}(d).
We observe that, as the interactions increases, around
$|\Delta_L/J_L|\approx 1+\sqrt{2}$ the contrast changes
from $\Con\approx 0$ to $\Con\approx 1$. 
It should be noted that in Fig.~\ref{fig:Fig1}(a-d) the chains are not very long and that there is a large interface tunnelling $J_M=J_L$ which broadens the transition.
In Fig.~\ref{fig:Fig1}(e,f) we show respectively the rectification $\Rec$ and the contrast $\Con$ as a function of the system size $N$ for different values of the anisotropy $\Delta_L/J_L$ (and for $J_M \ll J_L$). There is indeed a sharp contrast between $\Delta_L/J_L=1.5$ and $\Delta_L/J_L=3$; for the latter the rectification is orders of magnitude larger.

\begin{figure}
\includegraphics[width=\columnwidth]{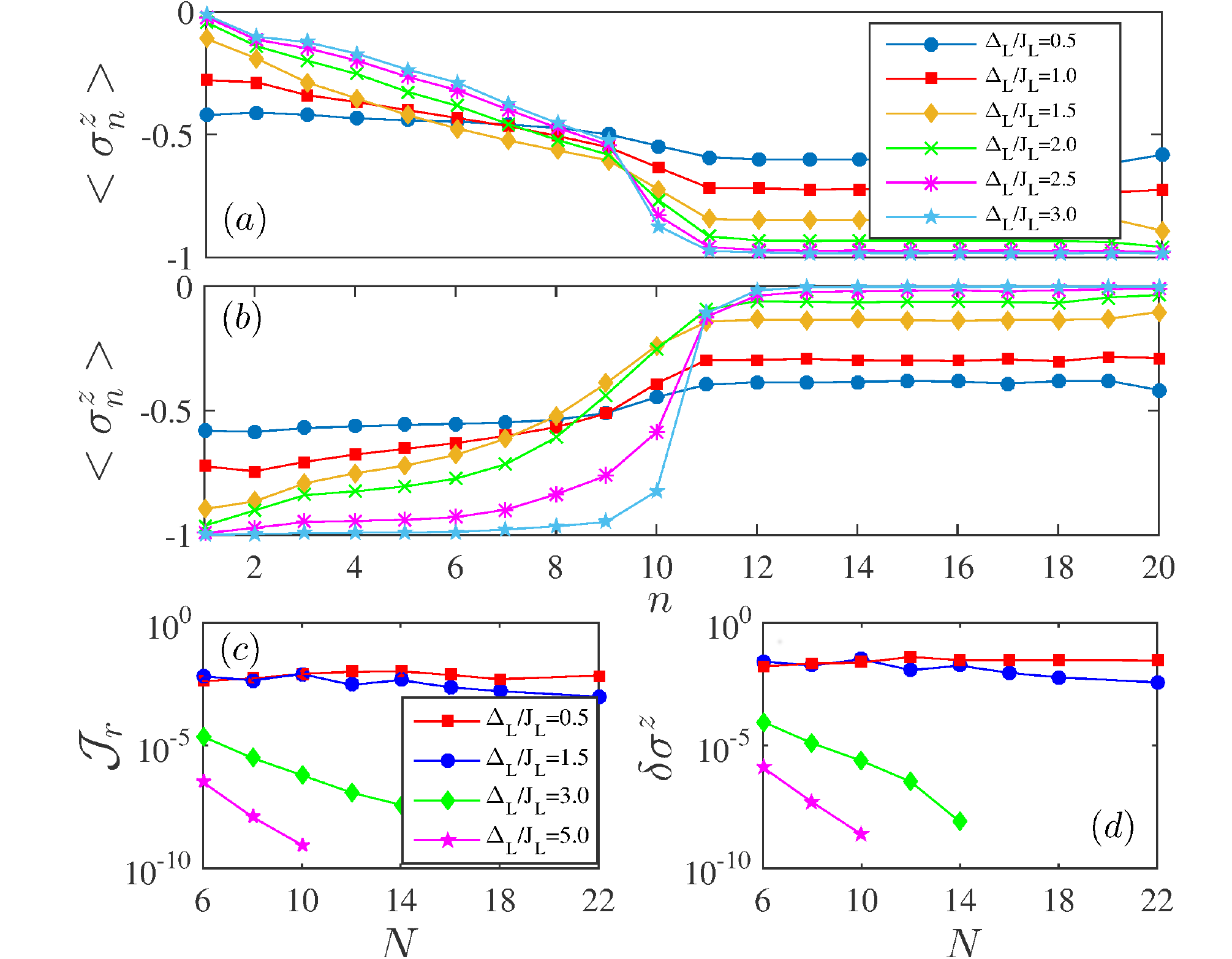}
\caption{(color online) Profile of the magnetization $\ave{\sop^z_n}$ as a function of the spin number $n$ for a chain of $N=20$ spins and for different values of the anisotropy $\Delta_L/J_L$:
(a) forward bias ($\la_1=0.5$ and $\la_N=0$) and (b) reverse bias ($\la_1=0$ and $\la_N=0.5$). In the bottom panels we show the spin current (c) and the magnetization difference $\delta\sigma^z$ (d) as a function of the system size in the reverse bias case. The other parameters values are as in Fig.~\ref{fig:Fig1}.}
\label{fig:Fig2}
\end{figure}

To understand the change in the transport properties of the system with the increase in anisotropy, we first study the magnetization profile, $\langle \sop^z_n\rangle$
as a function of the position $n$. In Fig.~\ref{fig:Fig2}(a) we focus on the forward bias case.
In the right half of the chain the magnetization is fairly constant, however, for large $\Delta_L/J_L$, a diffusive nature of transport is evidenced by the linear slope
of the magnetization in the left half of the chain.
For the reverse bias case the magnetization profile is very different, see Fig.~\ref{fig:Fig2}(b). In this case, for large anisotropy the change in magnetization becomes more marked, switching rapidly at the interface of the two half-chains between $-1$ and $0$.
Such large interface resistance implies a low spin current and suggests insulating behavior.

To confirm the insulating character of the system at large anisotropy and
in reverse bias, we study the spin current as a function of the system size $N$.
Fig.~\ref{fig:Fig2}(c) indeed indicates an exponential drop of the spin current
with $N$ at $\Delta_L/J_L=3$ and $5$, while at lower $\Delta_L/J_L$ the current hardly varies ($\Delta_L/J_L=0.5$) or decreases much more slowly ($\Delta_L/J_L=1.5$).
We also investigate the magnetization at site $n=1$.
This site is directly coupled to a bath that tries to set the magnetization to $-1$.
If the magnetization at site $1$ is exactly $-1$, then there would be
no effect of the bath on the chain, hence no transport
(note in fact that in the forward bias scenario, the magnetization of
the first and last spin does not match the one imposed by the baths).
Hence we study in Fig.~\ref{fig:Fig2}(d) the quantity $\delta\sigma^z=\ave{1+\sop^z_1}$,
which we refer to as the `magnetization difference', versus the length of
the spin chain, $N$, for different values of the anisotropy parameter.
Here we notice that for $\Delta_L/J_L=3$ and $5$ the magnetization difference decreases exponentially
as the system size increases, again a signature of an insulating regime, while, for smaller $\Delta_L/J_L$, the decrease is markedly slower.

\begin{figure}
\includegraphics[width=\columnwidth]{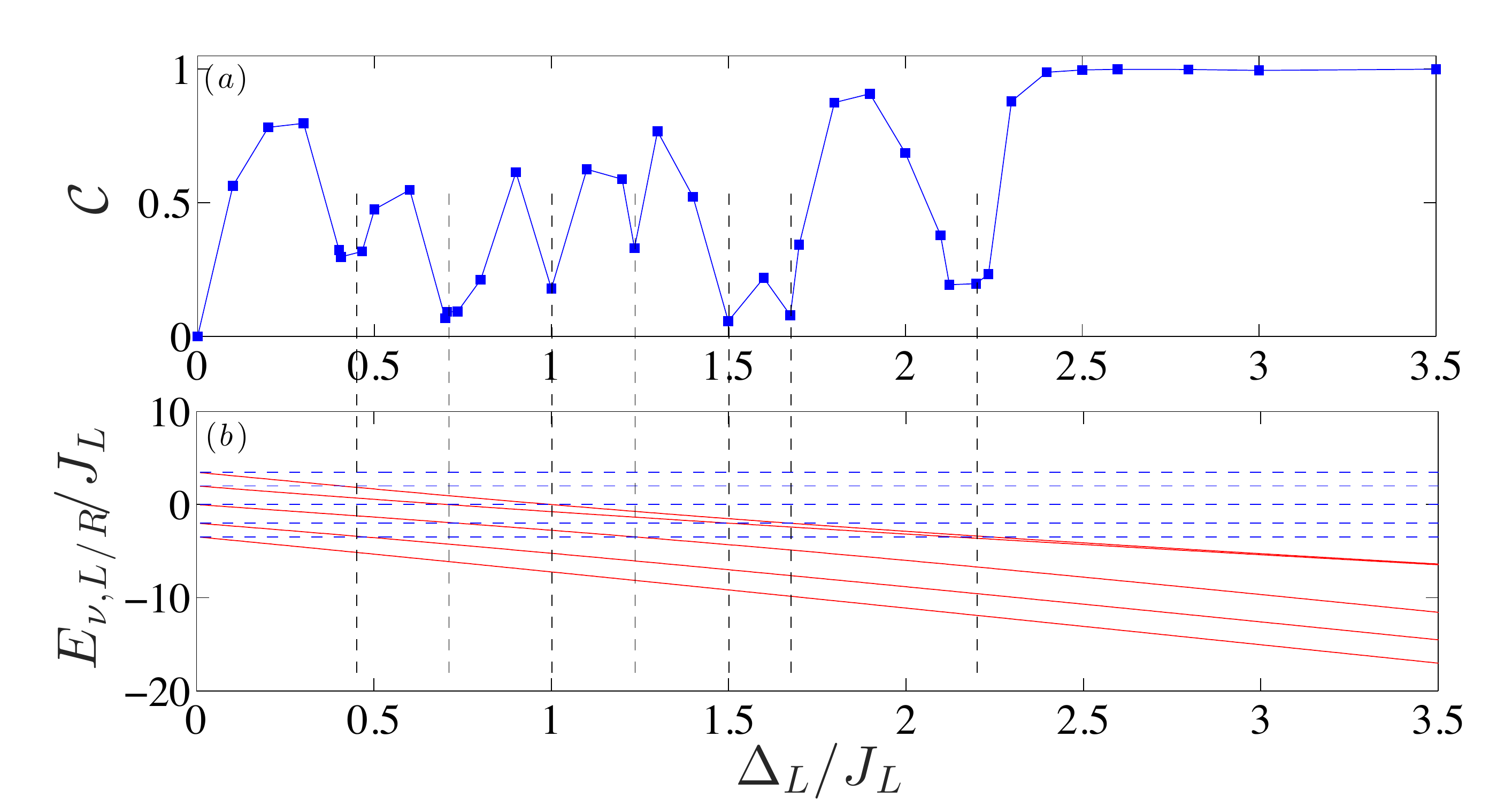}
\caption{(color online) (a) Contrast $\Con$ versus anisotropy $\Delta_L/J_L$.
(b) Eigenfrequencies $E_{\nu,R}$ and $E_{\nu,L}$ as a function of $\Delta_L/J_L$ computed from Eq.~(\ref{eq:matr}) for the right half (blue dashed lines) and left half of the chain (red continuous lines).
The black dashed lines show that the dips in the contrast $\Con$ in panel (a) occur when $E_{\nu,R}+E_{\nu,L}=0$. Parameter values: $N=10$, $J_M=J_L/10$, $\gamma=J_L/(10\hbar)$, $J_R=J_L$.}
\label{fig:Fig3}
\end{figure}

{\it The rectification mechanism}:
To explain the mechanism behind rectification
and highlight the role of interactions, we study a small chain with a weak coupling between the two halves as well as with the baths, i.e. $J_L=J_R$, $J_M=0.1 J_L$ and $\gamma=0.1J_L/\hbar$. In Fig.~\ref{fig:Fig3}(a) we find that the contrast $\Con$ versus the anisotropy has a resonant behavior, and, for certain values of $\Delta_L/J_L$, the contrast drops significantly, indicating almost no rectification; for $\Delta_L/J_L \gtrsim 1+\sqrt{2}$, the contrast is close to $1$, corresponding to strong rectification. Note that the use of small $\gamma$ and $J_M$ allows to clearly identify the resonances which was not possible for the parameters chosen in Fig.~\ref{fig:Fig1}(a-d).

The observed dips and the general rectification mechanism can be explained by the following considerations.
Let us take two chains: one fully polarized, i.e. $\rhop_{ss,\la_j=0}=\bigotimes_n \ket{\dw}_n\bra{\dw}$, while the other is at the infinite temperature state, i.e. $\rhop_{ss,\la_j=0.5}=\bigotimes_n \left(\ket{\uw}_n\bra{\uw}+\ket{\dw}_n\bra{\dw}\right)/2$. As soon as these chains are coupled together, any magnon excitation generated at the interface propagates, provided there is no energy gap in the set-up made by the two half-chains.
If instead there is a gap, the excitation is localized, and cannot propagate through the system. As a result the system becomes insulating. 
We consider first the case for which the half chain with non-zero anisotropy is polarized, i.e. $\rhop_{ss,\la_j=0}$ (reverse bias), and we compute the energy required to generate a localized excitation at the edge.  The magnon excitation spectrum for this half chain is given by the eigenvalues of the following Toeplitz bordered matrix \cite{eigenvalues}.
\begin{align}
\mathds{M}(\Delta)=\left[\begin{array}{cccccc}
 -2\Delta_L & 2J_L & 0 & \dots & \dots & \dots  \\
2J_L & -4\Delta_L & 2J_L & 0 & \dots & \dots \\
0 & 2J_L & -4\Delta_L & 2J_L & 0 & \dots \\
\dots & \dots& \dots& \dots& \dots& \dots \\
\dots &\dots & 0 & 2J_L & -4\Delta_L & 2J_L \\
\dots &\dots & 0 & 0 & 2J_L & -2\Delta_L 
\end{array}\right]. \label{eq:matr}
\end{align}
We plot these eigenvalues $E_{\nu,L}$ in Fig.~\ref{fig:Fig3}(b) as a function of $\Delta_L/J_L$ for a half chain of length $N/2=5$ with red continuous lines. For large enough anisotropy the eigenvalues are confined between $-2\Delta_L$ and $-4\Delta_L$.
This chain is coupled to another one with no anisotropy, $\Delta_R=0$, prepared in the state $\rhop_{ss,\la_j=0.5}$. The energy spectrum $E_{\nu,R}$ for a single excitation in this chain is also given by the eigenvalues of (\ref{eq:matr}) after exchanging $J_L$ with $J_R$ and setting the anisotropy to $0$. The energies $E_{\nu,R}$ are, for a large chain, within $\pm 4 J_R$ and for a half-chain of $N/2=5$ they are shown in Fig.~\ref{fig:Fig3}(b) with blue dashed lines.
The energetic cost of generating a magnon excitation is thus given by $E_{\nu,L}+E_{\nu',R}$. 
If $E_{\nu,L}+E_{\nu',R}\ne 0$ for all $\nu,\nu'$, then the generation of an excitation in the left and right halves of the chain requires to overcome an energy gap and transport will be hindered. The magnon excitation will be exponentially localized at the interface and will not be able to reach the baths.
In Fig.~\ref{fig:Fig3}(b) we use vertical black dashed lines to clearly show that when $E_{\nu,L}=-E_{\nu',R}$ (note that $E_{\nu',R}$ is symmetric around $0$) the contrast is significantly suppressed because of the absence of a gap.

In the thermodynamic limit for the reverse bias, the presence of a gap for the magnon excitation results in a marked change of transport properties. In fact, for longer chains, the spectrum for the half-chain without anisotropy approaches a continuum between $\pm 4J_R$, hence there will be no significant rectification as long as the largest of the $E_{\nu,L}$ is smaller than $-4 J_R$. This is why, for example, in Fig.~\ref{fig:Fig1}(c,d) the rectification and contrast are very small for low interactions. For large interactions, $\Delta_L/J_L>1$, and in the thermodynamic limit, the largest eigenvalue for the magnon excitation is $E_{{\rm max},L}=-2\Delta_L+2J_L^2/\Delta_L$ \cite{expsolution}. Hence, the critical anisotropy $\Delta_{L,c}$, at which the rectification changes significantly, becomes
\begin{align} \label{eq:critDelta}
\Delta_{L,c}=J_R+\sqrt{J_R^2+J_L^2}.
\end{align}
It is thus clear that for $J_L=J_R$, the critical anisotropy in the thermodynamic limit is $\Delta_{L,c}/J_L=1+\sqrt{2}$. This is consistent with the numerical results shown in Fig.~\ref{fig:Fig2}(c,d) which show that the current in reverse bias decreases exponentially with the system size. For completeness, in \cite{supp} we also show a characteristic time scale needed for a magnon excitation to propagate in the chain. Our numerics indicate that, for interaction strengths beyond the critical value $\Delta_{L,c}$, the time scale goes to infinity as the system size increases.

For the forward bias case instead, there will always be low energy excitations in the half-chain with non-zero anisotropy which can be matched by the quadratic chain, whose excitation spectrum does not change, resulting in good transport of excitations (for more details see \cite{supp}).
Hence there is a very different non-equilibrium response in forward and reverse bias.

\begin{figure}
\includegraphics[width=\columnwidth]{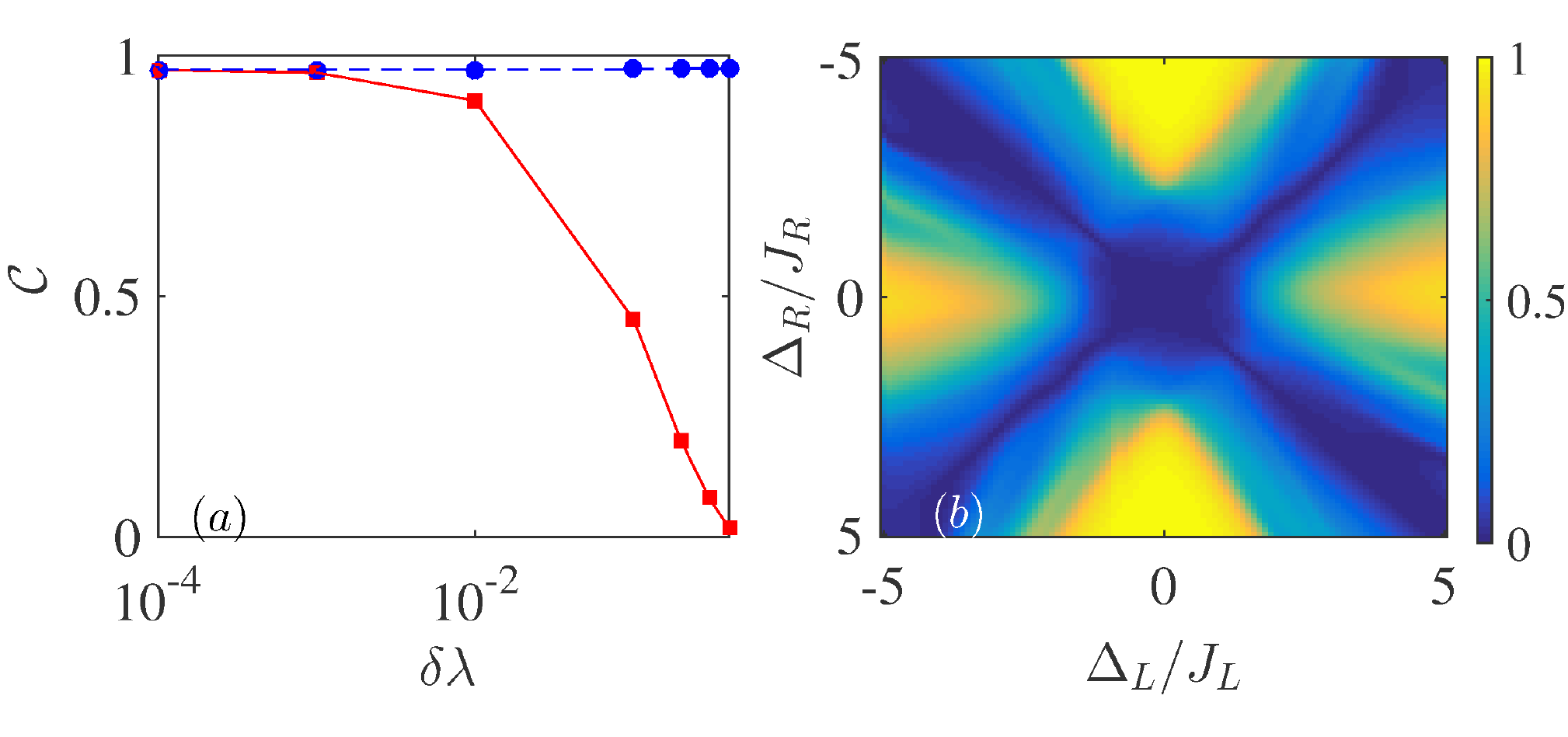}
\caption{(color online) (a) Contrast $\Con$ versus $\delta \lambda$ which is the distance from $\lambda_n=0$ while the other bath parameter is kept at $\lambda_{n'}=0.5$ (red continuous line) or the distance from $\lambda_n=0.5$ while the other bath parameter is kept at $\lambda_{n'}=0.0$ (blue dashed line). Parameters are: $N=10$, $\Delta_R=\Delta_M=0$, $\Delta_L/J_L=3$, $\gamma=J_R/\hbar$ and $J_M=J_L=J_R$. (b) Contrast $\Con$ as a function of both $\Delta_L/J_L$ and $\Delta_R/J_R$. Parameters are: $N=6$, $J_M=J_L=J_R$ and $\gamma=J_R/\hbar$.}
\label{fig:Fig4}
\end{figure}

{\it Robustness of the rectification}: It is important to study the robustness of the rectification to variations of the baths or Hamiltonian's parameters. First we consider the case for which the bath parameters $\lambda_n$ are not set exactly to either $0$ or $0.5$. The continuous blue line in Fig.~\ref{fig:Fig4}(a) shows the contrast $\Con$ when the parameter $\lambda_n$ which should be $0$ (i.e. $\lambda_N$ in forward bias and $\lambda_1$ in reverse bias) is detuned by an amount $\delta\lambda$, while the other is kept at $0.5$ (red continuous line). The contrast is close to $1$ as long as $\delta\lambda<10^{-2}$, while for $\delta\lambda=0.1$ the contrast is $\Con\approx 0.5$ corresponding to a rectification coefficient $\Rec \approx 3$.
We now consider instead the case when the maximum value of $\lambda_n$ is not $0.5$ but $0.5-\delta\lambda$ (while the other is kept at $0$). In this case the rectification is large and robust, as signalled by a contrast $\Con$ close to $1$ even at $\delta\lambda=0.1$ (blue dashed line). The rectification is thus very robust to changes of one of the bath parameters (the one that sets the local magnetization to $0$), while more care is needed in ensuring that one of the baths is almost completely polarized.

In Fig.~\ref{fig:Fig4}(b) we show how the contrast $\Con$ changes as we keep $J_L=J_R$ and vary both $\Delta_L$ and $\Delta_R$. The contrast is pronounced when the modulus of the anisotropies is different, with largest contrast when one of the two is $0$. This is expected because a larger excitation gap is present when one half of the chain is quadratic. The symmetric behavior is due to the spectrum invariance for a change of all $\Delta_n\rightarrow -\Delta_n$.

{\it Conclusions}: We have shown the possibility of large spin-current rectification even in short segmented spin chains.
In the thermodynamic limit,
our results are consistent with the prediction of an insulating behavior for reverse bias, leading to a perfect diode.
The key ingredient for such effect is the presence of a 
magnon excitation gap between the two chains, which is possible because of large enough anisotropy in one half of the system.
It is important to stress that in our set-up the rectification is maximum when half of the chain is quadratic (XX chain).

Materials whose spin dynamics is well described by Heisenberg chains with $\Delta_L/J_L=1.2$ or larger have been already studied \cite{Janson}. Such anisotropy can suffice to obtain large rectification if such material is matched to one with lower tunnelling parameter $J_R<J_L$ and negligible anisotropy ($\Delta_R=0$) as shown from Eq.~(\ref{eq:critDelta}). We also point out that recent proposals \cite{Lukin, Hannaford,Brune} and experiments with Rydberg atoms trapped with optical tweezers offer tremendous opportunities to engineer spin Hamiltonians with arbitrary values for the ratio $\Delta_{L}/J_{L }$ \cite{Lukin1, BarredoBrowaeys2016}. Moreover, cutting-edge technologies using adatoms on surface \cite{Otte} or trapped ions \cite{Cirac,Roos,Lewenstein} provide alternative frameworks for obtaining our proposed spin chain model.

The rectification, here demonstrated for spin currents, is also expected for heat currents imposed by heat baths, due to the mismatch in spectral response between the two portions of the chain.
More generally many-body interactions and phase transitions in the non-equilibrium scenarios are opening the door to the design of highly performing quantum devices, a research direction which should be further investigated.

{\it Acknowledgments}: We acknowledge fruitful discussions with C. Guo and T. Prosen. D.P. and V. B. acknowledge support from SUTD-MIT IDC (Project No. IDG31600107), and Singapore Ministry of Education, Singapore Academic Research Fund Tier-I (project SUTDT12015005). E.P. was partially supported by CNPq (Brazil). G.B. acknowledges the financial support of the INFN through the project “QUANTUM”. The computational work for this article was partially performed on resources of the National Supercomputing Centre, Singapore \cite{NSCC}.


\setcounter{equation}{0}
\setcounter{figure}{0}
\renewcommand{\theequation}{S\arabic{equation}}
\renewcommand{\thefigure}{S\arabic{figure}}
\renewcommand{\bibnumfmt}[1]{[S#1]}
\renewcommand{\citenumfont}[1]{S#1}

\section*{Supplementary material}
\subsection{Response in forward bias for the anisotropic case}
For the forward bias, the left-half of the chain with $\Delta_L\ne 0$ has a more complex response to a spin-flip at its edges. However we can show that there are low frequencies modes that can be matched with the right half of the chain. 
We consider the two-time correlator $S(t)=\langle\sigma^+_{N/2}(t)\sigma^-_{N/2}(0)\rangle_{ss}=\tr\left(\sigma^+_{N/2} e^{\Lop t} \sigma^-_{N/2}\;\rho_{ss}\right)$, and its Fourier transform $S(\omega)$, where $\hat{\rho}_{ss}$ is the steady state. We consider a small (negligible) dissipation. The frequency response of $S(\omega)$ is given by the differences of the energies $E_{\nu,L/R}$ occupied by the steady state and excited by $\sigma^-_{N/2}$. Note that the spectrum of right half of the chain is unaffected by the bias (e.g. whether $\lambda_j=0.5$ to $0$).  As for the steady state of left half of chain in forward bias, it is given by  $\hat{\rho}_{ss,\lambda_j=0.5}=\bigotimes_{n=1}^{N/2} \left(\ket{\uw}_n\bra{\uw}+\ket{\dw}_n\bra{\dw}\right)/2$.
The action of the operator $\sigma^-_{N/2}$ on the steady state results in $\Oop=\sigma^-_{N/2}\hat{\rho}_{ss,\lambda_j=0.5}=\left[\bigotimes_{n=1}^{N/2-1}\left(\ket{\uw}_n\bra{\uw}+\ket{\dw}_n\bra{\dw}\right)/2\right]\ket{\dw}_{N/2}\bra{\uw}$. Considering (i) the symmetry of the Hamiltonian to a global spin-flip, (ii) the fact that the Hamiltonian preserves the magnetization and (iii) focusing on the states near zero total magnetization (which are indeed present in $\Oop$), it is easy to see that for half-chains of odd length there is a non-zero response of $S(\omega)$ at $\omega=0$, while for even chains there be a finite response for a frequency which vanishes as the system size increases. For example let us consider $\zop=\ket{\uw \dw \dw}\bra{\uw \dw \uw}$ for a half-chain of length $N/2=3$, which is included in $\Oop$: For small dissipation, the Hamiltonian evolution, given by $d\zop/dt=\Lop(\zop)=-(\im/\hbar)\left[\Hop,\zop\right]$, connects $\zop$ to other states whose total number of spins up (or down) in the bra and in the ket remains unchanged (for example $\zop'=\ket{\dw \dw \uw}\bra{\dw \uw \uw}$). The evolution due to the anticommutation with the Hamiltonian is given by the difference of two phases. But since the Hamiltonian is symmetric to a global spin-flip, and since the total number of spins up minus spins down is opposite between the bra and the ket (in this example it is $-1$ for the ket and $+1$ for the bra), then there is a finite response for $S(\omega)$ at $\omega=0$.\\

\subsection{Propagation time in the reverse bias scenario}

\begin{figure}
\includegraphics[width=\columnwidth]{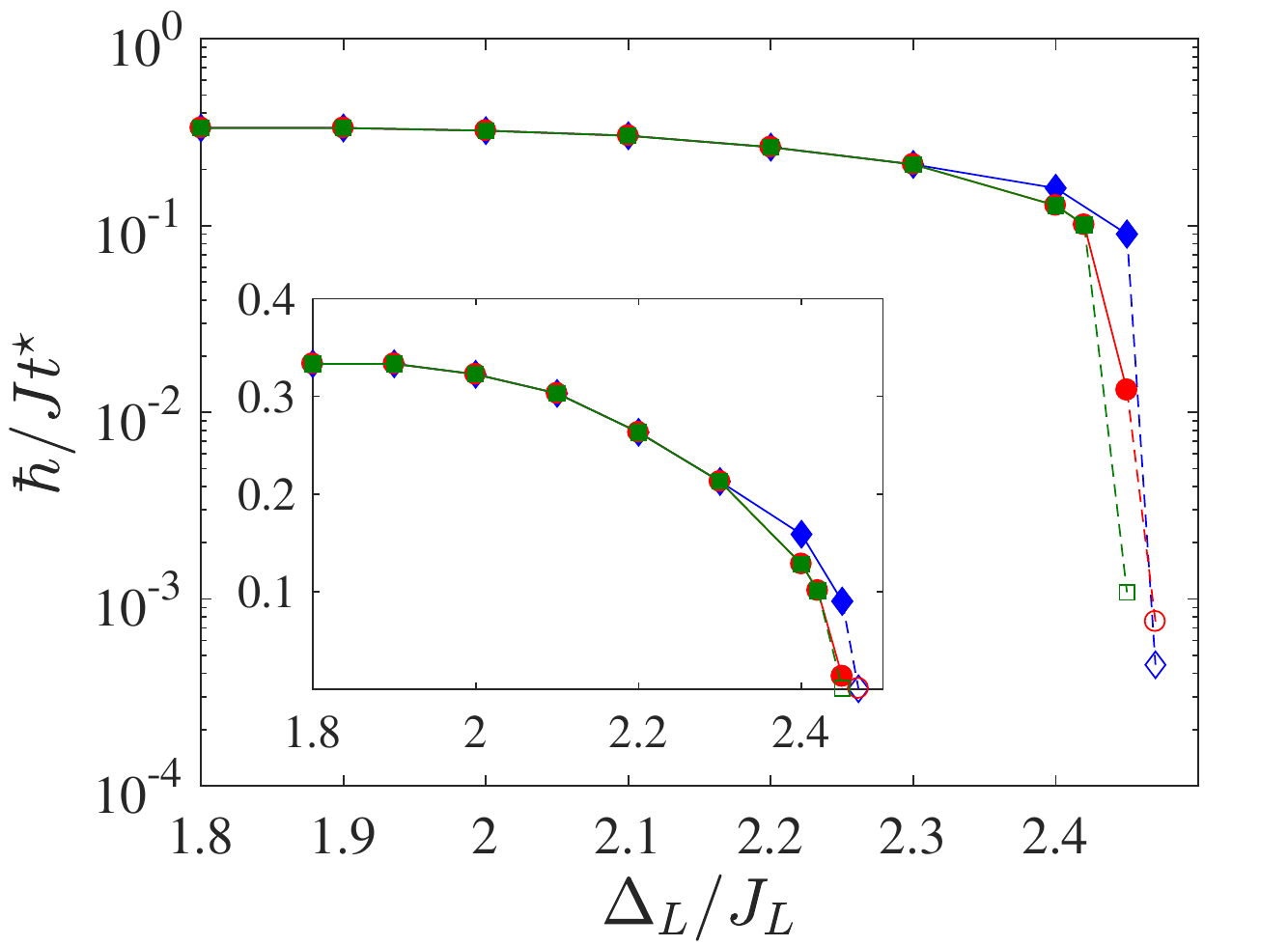}
\caption{(color online) Inverse propagation time $1/t^\star$ versus anisotropy $\Delta_L/J_L$ for different chain lengths, from top to bottom, $N=20$ (blue diamonds), $50$ (red circles), and $100$ (green squares). The empty symbols represent a numerical lower bound to $t^\star$. The inset shows a lin-lin plot while the main figure depicts the log-lin plot. Parameters are: $J_M=0.05J_R$ and $J_L=J_R$.}
\label{fig:Fig1S}
\end{figure}

Here we give another indication of the presence of a transition to an insulating regime in the reverse bias scenario. Let us consider two chains each of length $N/2$: an $XXZ$ chain (with anisotropy $\Delta_L$ and tunnelling $J_L$) prepared with spins all pointing down, and an $XX$ chain ($J_R=J_L$) prepared in the infinite temperature state. We then connect these chains with an $XX-$type of coupling with small magnitude $J_M$ and study the ensuing dynamics. Note that here no baths are used, but we just study Hamiltonian dynamics. We focus in particular on the spin of the $XXZ$ chain closest to the interface between the two chains. At time $t=0$ this spin is pointing down, i.e. $\langle \sop^z_{N/2}\rangle=-1$ and if the chain is not insulating, its polarization will change. We then compute the propagation time $t^\star$ at which $\langle \sop^z_{N/2}(t^*)\rangle=-1+\epsilon$ where $\epsilon$ is a small positive number which we take here to be $10^{-2}$. In Fig.~\ref{fig:Fig1S} we show the inverse of the propagation time $1/t^\star$ as a function of the interaction $\Delta_L$ for various chain lengths, each indicated by a different symbol. Note that for the empty symbols the magnetization at site $N/2$ has not yet reached the value $1-\epsilon$ despite the long time evolution (of the order of $1000 J/\hbar$); Hence we just plot the longest simulated time, which is a numerical lower bound to $t^\star$. We see clearly that as the size of the system increases, the inverse of the propagation time goes quickly to $0$ as soon as the anisotropy exceeds a critical value close to $J_L(1+\sqrt{2})$ as predicted in the main article.

\end{document}